\documentclass[12pt]{article}
\usepackage[english]{babel}
\usepackage{amsmath}
\usepackage{bbm}
\usepackage{multirow}
\usepackage{amssymb}
\usepackage{bbold}
\usepackage{color}
\usepackage[titletoc,toc,title]{appendix}
\Roman{section}
\renewcommand{\Re}{\mathrm{Re}}
\renewcommand{\Im}{\mathrm{Im}}

\newcommand{\eq}[1]{\begin{equation}#1\end{equation}}
\newcommand{\naw}[1]{\left(#1\right)}
\newcommand{\ket}[1]{\left|#1\right>}
\newcommand{\bra}[1]{\left<#1\right|}
\newcommand{\av}[1]{\left<#1\right>}

\newcommand{\modu}[1]{\left|#1\right|}
\newcommand{\poisson}[1]{\left\{#1\right\}}

\begin{document}
\begin{center}
\textsc{\Large{On Nash equilibria in Eisert-Lewenstein-Wilkens game}}
\newline

\large{Katarzyna Bolonek-Laso\'n}\footnote{kbolonek@uni.lodz.pl}\\ 
\emph{\normalsize{Faculty of Economics and Sociology, Department of Statistical Methods, \\University of Lodz,
41/43 Rewolucji 1905 St., 90-214 Lodz,  Poland.}}\\
\large{Piotr Kosi\'nski}\footnote{pkosinsk@uni.lodz.pl}\\
\emph{\normalsize{Department of Computer Science, Faculty of Physics and Applied Informatics, University of Lodz, 149/153 Pomorska St., 90-236 Lodz, Poland.}}
\end{center}

\begin{abstract}
Landsburg method of classifying mixed Nash equilibria for maximally entangled Eisert-Lewenstein-Wilkens (ELW) game is analyzed with special emphasis on symmetries inherent to the problem. Nash equilibria for the original ELW game are determined. 
\end{abstract}

\section{Introduction}
The papers of Eisert, Lewenstein and Wilkens (ELW) \cite{EisertWL}, \cite{EisertW} and Meyer \cite{Meyer} opened new field of research - the theory of quantum games \cite{Marinatto}$\div$\cite{Bolonek5}. Roughly speaking this theory deals with the properties of the games obtained from classical ones by admitting more general correlations, in particular those which break Bell-type inequalities.

One of the most important notions of game theory is that of Nash equilibrium \cite{Nash}. The famous Nash theorem states that under quite general assumptions the classical noncooperative game exhibits Nash equilibrium (or equilibria). Its validity extends to quantum domain \cite{Lee}, \cite{Glicksberg}.

ELW game constructed in Ref. \cite{EisertWL} provides a paradigm of quantum game. The natural question arises concerning the description and classification of its Nash equilibria. The most interesting case seems to be that of maximal entanglement when the initial state of the game is maximally entangled. In this case some examples of Nash equilibria were presented in Ref. \cite{EisertW}. In a series of nice papers \cite{Landsburg}, \cite{Landsburg4}$\div$\cite{Landsburg3} Landsburg described the general structure of possible, in general mixed, Nash equilibria for maximally entangled ELW games. His classification is slightly involved because the problem is set in full generality, i.e. for arbitrary classical payoff matrix. 

Using the isomorphism between the group $SU(2)$ and the set of unit quaternions one can reformulate the ELW game in terms of quaternion algebra. This formalism takes a particularly simple form in the case of maximal entanglement and is extensively used in Landsburg's papers.\\
On the other hand it has been shown \cite{Bolonek} that the maximally entangled ELW game is also distinguished by the existence of the real structure in the Hilbert space of the states of the game which fully determine its properties. Therefore, equivalently the game can be analyzed in terms of geometry of real fourdimensional arithmetic space.

In the present paper we classify all Nash equilibria for the original ELW game, i.e. for the particular payoff matrix used in Ref. \cite{EisertWL}. We follow Landsburg's method with some modifications and special emphasis put on symmetries inherent in the problem. Landsburg analysis is based on two pillars. First, it is shown that any mixed strategy is equivalent to the one supported on at most four orthogonal quaternions/real fourvectors; second, the analysis of possible mixed strategies reduces to that of degeneracies of highest eigenvalues of the matrices determining the players payoffs. These two principles when combined with the symmetries of the problem give a very transparent picture.

The paper is organized as follows. I Sec. II we present the ELW game; then, in Sec. III the quaternionic formalism is introduced. Symmetries of ELW games, in particular their maximally entangled version, are discussed in Sec. IV. In the next section the Nash equilibria are introduced and Landsburg method described in more detail. In Sec. VI the Nash equilibria of the original ELW game are classified.

\section{The Eisert-Lewenstein-Wilkens quantum games (ELW games)}
Eisert, Lewenstein and Wilkens \cite{EisertWL}, \cite{EisertW} proposed a quantization scheme of the simplest classical symmetric noncooperative game involving two players, each having two classical strategies at his/her disposal. Let us denote the players by A (Alice) and B (Bob) while the relevant strategies are $C$ (cooperate) and D (defect). The classical game is completely determined by the payoff matrix

\eq{
\begin{tabular}{c|c|c|c|}
\multicolumn{2}{c}{} & \multicolumn{2}{c}{B}\\
\cline{3-4}
\multicolumn{1}{c}{} & & C & D\\
\cline{2-4}
\multirow{2}{1cm}{A} & C & $\naw{X^0,X^0}$ & $\naw{X^2,X^1}$\\
\cline{2-4}
 & D & $\naw{X^1,X^2}$ & $\naw{X^3,X^3}$\\
\cline{2-4}
\end{tabular}
}
where $X_\alpha$'s are the relevant outcomes. We will be dealing with the games which obey the conditions: (i) all $X^\alpha$ are distinct; (ii) all twofold sums $X^\alpha+X^\beta$ are distinct as well; such games are called generic in Landsburg terminology \cite{Landsburg}, \cite{Landsburg4}$\div$\cite{Landsburg3}. The properties of both classical and quantum games depend strongly on the actual values of the outcomes. In particular, the ordering 
\eq{X^1>X^0>X^3>X^2}
leads to the famous Prisoner Dilemma \cite{Myerson}. Depending on the context the additional constraints can emerge. For example, one can add \cite{DuLi3}
\eq{2X^0>X^1+X^2}
which implies  that in the iterated game the players are at least as well of playing always $(C,C)$ or alternating between $\naw{C,D}$ and $\naw{D,C}$.

In order to construct the quantum counterpart of classical game we ascribe to each player a twodimensional complex Hilbert space $H$; the total space of the game is the tensor product $H\otimes H$. The basic vectors in $H$ correspond to two classical strategies
\eq{\ket{C}=\left(\begin{array}{c}
1\\
0
\end{array}\right),\qquad \ket{D}=\left(\begin{array}{c}
0\\
1
\end{array}\right).\label{ab5}}
The initial state of the game is given by
\eq{\ket{\Psi_i}\equiv J\naw{\ket{C}\otimes \ket{C}}\label{ab}}
where $J$ is the gate operator which introduces the quantum entanglement leading to quantum correlations breaking, in general, Bell-type inequalities.  The form of $J$ is determined by  two conditions
\begin{itemize}
\item[(i)] the quantum game continues to be symmetric with respect to the exchange of players;
\item[(ii)] the classical strategies are properly represented in the quantum game.
\end{itemize}
These conditions lead to the following form of gate operator \cite{EisertWL}
\eq{J=exp\naw{-\frac{i\gamma}{2}\sigma_2\otimes\sigma_2}\label{ab1}}
with $\gamma\in\av{0,\frac{\pi}{2}}$ and $\sigma_2$ being the Pauli matrix. One obtains a one parameter family of gate operators. Quantum strategies of Alice and Bob are represented by unitary matrices $U_A$ and $U_B$ belonging to $SU(2)$. The final state of the game is defined as
\eq{\ket{\Psi_f}=J^+\naw{U_A\otimes U_B}J\naw{\ket{C}\otimes\ket{C}}}
and allows us to compute Alice and Bob payoffs
\eq{
\begin{split}
& \$_A=X^0\modu{\naw{\bra{C}\otimes\bra{C}}\ket{\Psi_f}}^2+X^1\modu{\naw{\bra{D}\otimes\bra{C}}\ket{\Psi_f}}^2+\\
&\qquad X^2\modu{\naw{\bra{C}\otimes\bra{D}}\ket{\Psi_f}}^2+X^3\modu{\naw{\bra{D}\otimes\bra{D}}\ket{\Psi_f}}^2
\end{split}\label{ab6}}
\eq{
\begin{split}
& \$_B=X^0\modu{\naw{\bra{C}\otimes\bra{C}}\ket{\Psi_f}}^2+X^1\modu{\naw{\bra{C}\otimes\bra{D}}\ket{\Psi_f}}^2+\\
&\qquad X^2\modu{\naw{\bra{D}\otimes\bra{C}}\ket{\Psi_f}}^2+X^3\modu{\naw{\bra{D}\otimes\bra{D}}\ket{\Psi_f}}^2.
\end{split}\label{ab7}}
The properties of the game depend on the choice of payoff matrix, the value of the parameter $\gamma$ and the choice of the  manifold $S\subset SU(2)$ of allowed strategies.  In the original ELW paper $S$ is some proper subset of $SU(2)$ which itself is not a group. It seems difficult to find a justification for the use of such set of strategies \cite{BenjaminHay}. Therefore, in what follows we assume that $S=SU(2)$, i.e. all elements of $SU(2)$ group are admissible as Alice and Bob strategies.

The value of $\gamma$ is also significant. For $\gamma=0$ we obtain the classical game. The set of all pure quantum strategies coincides then with the set of all classical mixed ones (some parameters of $SU(2)$ group become redundant). On the other hand, $\gamma=\frac{\pi}{2}$ yields maximal entanglement of initial state; the  properties of the game change radically. For example, as we shall see, due to the Bell-like correlations, the quantum game lacks nontrivial pure Nash equilibria.

The payoffs of the players depend on the actual values of $X^\alpha$'s. However, some outcomes coincide due to the purely group-theoretical reasons. To see this let us define the stability subgroup $G_s\subset SU(2)\times SU(2)$ of the initial state:
\eq{g\in G_s\Rightarrow g\ket{\Psi_i}=\ket{\Psi_i}.}
Using eqs. (\ref{ab}) and (\ref{ab1}) it is easy to determine $G_s$. We conclude that $G_s$ depends on the value of $\gamma$. For $\gamma\neq \frac{\pi}{2}$ one obtains
\eq{G_s=U(1)}
and $g\in G_s$ has the form 
\eq{g=e^{i\varphi\frac{\sigma_3}{2}}\otimes e^{-i\varphi\frac{\sigma_3}{2}}.}
In the case of maximal entanglement, $\gamma=\frac{\pi}{2}$, $G_s$ becomes, up to an isomorphism, the diagonal subgroup of $SU(2)\times SU(2)$. More precisely, any $g\in G_s$ has the form 
\eq{g=\naw{U,\overline{U}},\quad U\in SU(2)}
i.e. $G_s\sim SU(2)$.\\
Obviously, two pairs of strategies, $\naw{U_A,U_B}$ and  $\naw{U_A',U_B'}$, differing only by an element of stability subgroup,
\eq{\naw{U_A,U_B}=\naw{U_A',U_B'}\cdot g\label{ab2}}
lead to the same outcome.

We see that the maximal entanglement corresponds to the largest stability group. This fact has profound consequences for the structure of maximally entangled game \cite{Bolonek5}. To see this note the identity in $SU(2)\times SU(2)$
\eq{\naw{U_A,U_B}=\naw{U_AU_B^T,I}\cdot\naw{\overline{U}_B,U_B}.\label{ab3}}
Comparying eqs. (\ref{ab3}) and (\ref{ab2}) we conclude that the outcomes of both players depend only on the product $U_AU_B^T$.

\section{The quaternionic formalism}
It is well known that the $SU(2)$ group is isomorphic to the group of unit quaternions. It is, therefore, not surprising that one can formulate the ELW scheme in terms of quaternion algebra. 

Let us remind shortly the notion of quaternions. One considers the fourdimensional real vector space spanned by the vectors $e_\alpha$, $\alpha=0,1,2,3$ which may be take as orthonormal ones, $\naw{e_\alpha,e_\beta}=\delta_{\alpha,\beta}$. The multiplication law is defined as follows
\eq{e_0=\mathbb{1},\quad e_i^2=-\mathbb{1},\quad e_ie_j=\varepsilon_{ijk}e_k,\quad i,j,k=1,2,3}
extended by linearity to any pair of vectors.\\ Putting
\eq{p=p_\alpha e_\alpha,\quad q=q_{\alpha}e_\alpha}
(summation over $\alpha$ understood) one easily finds
\eq{\begin{split}
& \naw{pq}_0=p_0q_0-p_iq_i\\
& \naw{pq}_i=p_0q_i+q_0p_i+\varepsilon_{ijk}p_jq_k
\end{split}}
(summations over latin indices run from 1 to 3).
Quaternions form a noncommutative field with conjugation 
\eq{\overline{p}\equiv p_0e_0-p_ie_i}
and quaternionic norm $\modu{p}$:
\eq{\modu{p}^2e_0=p\overline{p}=\naw{\sum_{\alpha=0}^3p_\alpha^2}e_0.}
The quaternion is called a unit one iff $\modu{p}=1$.

 Any matrix $U\in SU(2)$ can be written as
\eq{U=p_0\mathbb{1}-ip_k\sigma_k,\quad p_0^2+\sum_{k=1}^3p_k^2=1.}
The mapping
\eq{U\longleftrightarrow p=p_\alpha e_\alpha\label{ab4}}
defines the group isomorphism. In particular, $e_k\leftrightarrow -i\sigma_k$, $k=1,2,3$ while $e_0$ is $2\times 2$ unit matrix.\\
Let us consider an ELW game defined by some parameter $\gamma\in\av{0,\frac{\pi}{2}}$. Using the isomorphism (\ref{ab4}) one can ascribe the quaternions $p$ and $q$ to the strategies of Alice and Bob, respectively,
\eq{U_A\longleftrightarrow p=p_\alpha e_\alpha\label{ab8}}
\eq{U_B\longleftrightarrow q=q_\alpha e_\alpha.}
The gate operator $J$, defined by eq. (\ref{ab1}) takes the following quaternionic form
\eq{J=\cos\naw{\frac{\gamma}{2}}e_0\otimes e_0+i\sin\naw{\frac{\gamma}{2}}e_2\otimes e_2.}
Finally, note that if we put
\eq{U\longleftrightarrow r=r_\alpha e_\alpha}
then, according to the definitions (\ref{ab5})
\eq{\bra{C}U\ket{C}=r_0-ir_3}
\eq{\bra{D}U\ket{C}=-ir_1+r_2.\label{ab9}}
Eqs. (\ref{ab6}), (\ref{ab7}) and (\ref{ab8})$\div$(\ref{ab9}) allow us to write the expected payoffs of Alice and Bob in terms of quaternions. This construction does not seem to be very useful except the case of maximal entanglement, $\gamma=\frac{\pi}{2}$. In the latter case the quaternionic form may be described as follows. Let
\eq{ U_A=\left( \begin{array}{cc}
a & b\\
-\overline{b} & \overline{a}\end{array}\right)\,\qquad \modu{a}^2+\modu{b}^2=1}
\eq{ U_B=\left( \begin{array}{cc}
\alpha & \beta\\
-\overline{\beta} & \overline{\alpha}\end{array}\right)\,\qquad \modu{\alpha}^2+\modu{\beta}^2=1}
be the strategies of Alice and Bob, respectively. The Alice strategy is represented by the quaternion 
\eq{p=\Re a\,e_0-\Im b\,e_1-\Re b\,e_2-\Im a\,e_3\label{z}}
which corresponds to the isomorphism (\ref{ab4}). On the other case, the quaternion corresponding to Bob strategy reads
\eq{q=\Re \alpha \,e_0+\Re \beta\, e_1+\Im \beta\, e_2+\Im \alpha\, e_3.\label{z1}}
This correspondence defines an isomorphism between $SU(2)$ and unit quaternions such that
\eq{e_1\longleftrightarrow i\sigma_2,\quad e_2\longleftrightarrow i\sigma_1,\quad e_3\longleftrightarrow i\sigma_3.}
Using the above definitions it is not difficult to check that the payoff functions $\$_A$ and $\$_B$ (eqs. (\ref{ab6}) and (\ref{ab7})) take the form
\eq{
\begin{split}
&\$_A\naw{p,q}=X_0\naw{pq^{-1}}_0^2+X_1\naw{pq^{-1}}_1^2+X_2\naw{pq^{-1}}_2^2+X_3\naw{pq^{-1}}_3^2\equiv\\& \qquad\qquad\equiv \sum_{\alpha=0}^{3}X_\alpha\naw{pq^{-1}}_\alpha^2\end{split}\label{ab10}}
\eq{
\begin{split}
&\$_B\naw{p,q}=X_0\naw{pq^{-1}}_0^2+X_2\naw{pq^{-1}}_1^2+X_1\naw{pq^{-1}}_2^2+X_3\naw{pq^{-1}}_3^2\equiv\\& \qquad\qquad\equiv \sum_{\alpha=0}^{3}\widetilde{X}_\alpha\naw{pq^{-1}}_\alpha^2\end{split}\label{ab11}}
with $\widetilde{X}_{0,3}=X_{0,3}$, $\widetilde{X}_{1,2}=X_{2,1}$.\\
We see that the payoff functions take a particularly simple form in quaternionic formalism. As expected, due to the large stability group the payoffs depend only on the product $pq^{-1}$. In what follows we make the replacement $q^{-1}\rightarrow q$ which simplifies eqs. (\ref{ab10}) and (\ref{ab11}) and amounts only to relabelling the Bob strategies.

\section{Symmetries of ELW games}
The ELW games exhibit a number of symmetries which appear to be more or less useful when considering their specific properties. First, the structure of the game (for example, the set of Nash equilibria) does not change, except the actual values of the payoffs, if one makes the substitution
\eq{X_{\alpha}\rightarrow\lambda X_\alpha+\mu,\quad\mu\in\mathbb{R},\quad \lambda\in\mathbb{R}_+ .}
Assume now that we are dealing with maximally entangled game. Let $r$ be any unit quaternion. Then eqs. (\ref{ab10}) and (\ref{ab11}) imply (with $q$ replaced by $q^{-1}$ as explained above)
\eq{\$_{A,B}\naw{p,q}=\$_{A,B}\naw{pr^{-1},rq}. }
In particular,
\eq{\$_{A,B}\naw{p,q}=\$_{A,B}\naw{\mathbb{1},pq}=\$_{A,B}\naw{pq,\mathbb{1}}.}
This property results, as it has been explained above, from the structure of stability group in maximally entangled case. 

 Another symmetry property is related to the specific geometry of quaternions \cite{Landsburg4}. It is well known that for unit quaternions $p$ and $q$ the transformation
\eq{r\rightarrow prq^{-1}\label{c}}
describes $SO(4)$ rotation of a four-vector $r$; in fact, eq. (\ref{c}) defines the local isomorphism $SU(2)\times SU(2)\sim SO(4)$. Let $\sigma \in S_4$ be any permutation of $0,1,2,3$. Then one can always adjust the signs $\pm$ of $e_{\sigma^{-1}\naw{\alpha}}$ such that
\eq{\pm e_{\sigma^{-1}\naw{\alpha}}=p_1e_\alpha q_1^{-1}}
for some unit quaternions $p_1$ and $q_1$.\\
Therefore, for any quaternion $r$
\eq{p_1rq_1^{-1}=\naw{\pm r_{\sigma\naw{\alpha}}}e_\alpha.}
As a result one can rewrite the expected payoff of (say) Alice as 
\eq{\$_A=\sum_{\alpha=0}^{3}X_\alpha\naw{pq}_\alpha^2=\sum_{\alpha=0}^{3}X_\alpha\naw{p_1^{-1}pqq_1}_{\sigma\naw{\alpha}}^2}
and the same holds true for the Bob payoff $\$_B$. We conclude that by permuting the components $\naw{pq^{-1}}_\alpha^2$ in eqs. (\ref{ab10}) and (\ref{ab11}) we obtain the equivalent games; they differ only from the initial one by relabelling the strategies only.

Both classical and quantum  games are symmetric with respect to the exchange of players. This symmetry is described on the quaternionic level as follows.  Let
\eq{r\equiv\frac{1}{\sqrt{2}}\naw{e_0+e_3};}
then
\eq{re_0r^{-1}=e_0,\quad re_1r^{-1}=e_2,\quad re_2r^{-1}=-e_1,\quad re_3r^{-1}=e_3.}
Noting that $\naw{pq}_\alpha^2=\naw{\overline{pq}}_\alpha^2=\naw{\overline{q}\cdot\overline{p}}_\alpha^2$ one finds
\eq{\$_A\naw{p,q}\equiv\$\naw{p,q}}
\eq{\$_B\naw{p,q}=\$\naw{r\,\overline{q}\,r^{-1},r\,\overline{p}\,r^{-1}}}
which expresses the symmetric role of both players.

\section{Mixed strategies and Nash equilibria}
The set of all pure strategies of each player consists of unit quaternions, i.e. it forms the unit threedimensional sphere $S_3$. Therefore, a mixed strategy is represented by a normalized  nonnegative measure $\mu$ on $S^3$. In general, both players are playing mixed strategies. Then the expected payoffs read
\eq{\$_A\naw{\mu,\nu}=\iint\limits_{S^3\times S^3}\$_A\naw{p,q}\mathrm{d}\mu\naw{p}\mathrm{d}\nu\naw{q}}
\eq{ \$_B\naw{\mu,\nu}=\iint\limits_{S^3\times S^3}\$_B\naw{p,q}\mathrm{d}\mu\naw{p}\mathrm{d}\nu\naw{q}.}
One of the most important notions in game theory is that of Nash equilibrium. Let us remind that a pair of strategies $\naw{\mu_0,\nu_0}$ defines a Nash equilibrium iff
\eq{\$_A\naw{\mu_0,\nu_0}\geq\$_A\naw{\mu,\nu_0}}
\eq{\$_B\naw{\mu_0,\nu_0}\geq\$_B\naw{\mu_0,\nu}}
for all strategies $\mu$ and $\nu$.\\
The problem of finding and classifying all Nash equilibria for general ELW game is nontrivial. The most interesting case seems to be the one corresponding to the maximally entangled game where the role of quntum correlations is the most significant. On the other hand the problem simplifies then due to the large symmetry of the game. In a series of papers \cite{Landsburg}, \cite{Landsburg4}$\div$\cite{Landsburg3}, Landsburg was able to classify all potential candidates for Nash equilibria. Their general description is still quite involved but the underlying basic ideas are nice and transparent. In fact, there are two basic steps to be performed. First, one has to find the "canonical" form of mixed strategy. To this end we start with the following definitions. We call two measures on $S_3$, $\nu$ and $\nu'$, left equivalent iff
\eq{\int\limits_{S^3}\naw{pq}_\alpha^2\textrm{d}\nu\naw{q}=\int\limits_{S^3}\naw{pq}_\alpha^2\textrm{d}\nu'\naw{q}}
for any unit quaternion  (and, consequently, for any quaternion) $p$ and $\alpha=0,1,2,3$. Analogously, two measures, $\mu$ and $\mu'$, are right equivalent iff
\eq{\int\limits_{S^3}\naw{pq}_\alpha^2\textrm{d}\mu\naw{p}=\int\limits_{S^3}\naw{pq}_\alpha^2\textrm{d}\mu'\naw{p}}
for any unit quaternion $q$ and $\alpha=0,1,2,3$.\\
It easy to classify all measures up to equivalence. To see this let us write
\eq{\naw{pq}_\alpha=m_{\alpha\beta}\naw{p}q_\beta.\label{c1}}
The matrix $m\naw{p}$ is orthogonal (if fact, $m\naw{p}\in SO(4)$ because $p\rightarrow\mathbb{1}\cdot p\cdot q$ is an $SO(4)$ transformation).
\eq{m\naw{p}m^T\naw{p}=I}
 It is easy to write out $m\naw{p}$ explicitly,
\eq{m\naw{p}=\left(\begin{array}{cccc}
p_0 & -p_1 & -p_2 & -p_3\\
p_1 & p_0 & -p_3 & p_2\\
p_2 & p_3 & p_0 & -p_1\\
p_3 & -p_2 & p_1 & p_0
\end{array}\right).}
Eq. (\ref{c1}) implies now
\eq{\int\limits_{S^3}\naw{pq}_\alpha^2\textrm{d}\nu\naw{q}=m_{\underline{\alpha}\beta}(p)m_{\underline{\alpha}\gamma}(p)\int\limits_{S^3}q_\beta q_\gamma\textrm{d}\nu(q)\label{c2}}
 (no summation over $\alpha$). The matrix
\eq{S_{\alpha\beta}\equiv\int q_\alpha q_\beta \textrm{d}\nu(q)}
is (a) real and symmetric; (b) positive semidefinite; (c) of unit trace. Therefore, it can be diagonalized by a real orthogonal transformation. In operator language, $S=\sum\limits_{a=1}^4\rho_a\ket{q^{(a)}}\bra{q^{(a)}}$, $\rho_a\geq 0$, $\sum\limits_{a=1}^4\rho_a=1$. Explicitly
\eq{S_{\alpha\beta}=\sum_{a=1}^4\rho_aq_\alpha^{(a)}q_\beta^{(a)}\label{e}}
and $q^{(a)}$ are orthonormal.
Let us note that the representation (\ref{e}) is not unique. If some eigenvalue $\rho_a$ is d-fold degenerate one can replace the corresponding eigenvectors $q^{(a1)},\, q^{(a2)},\ldots,\, q^{(ad)}$ by their arbitrary $O(d)$ combinations. In particular, if all $\rho_a$ are nondegenerate we are left with sign arbitrariness, $q^{(a)}\rightarrow\pm q^{(a)}$.
Then eq. (\ref{c2}) takes the form 
\eq{\int\limits_{S^3}\naw{pq}_\alpha^2\textrm{d}\nu(q)=\sum_{a=1}^4\rho_am_{\underline{\alpha}\beta}\naw{p}m_{\underline{\alpha}\gamma}\naw{p}\int q_\beta^{(a)}q_\gamma^{(a)}\text{d}\nu(q)=\sum_{a=1}^4\rho_a\int\naw{pq^{(a)}}_\alpha^2\text{d}\nu(q).\label{c3}}
Eq. (\ref{c3}) shows that any measure $\nu$ is left equivalent to the one supported on at most four orthonormal quaternions \cite{Landsburg}. The same holds true for right equivalence.

From the above definitions it is obvious that in computing the payoffs $\$_A$ and $\$_B$ one can replace Alice (Bob) measure by right (left) equivalent one. We conclude that the payoff functions can be always written in the form
\eq{
 \$_{A,B}\naw{\mu,\nu}=\sum_{a,b=1}^4\sigma_a\rho_b\$_{A,B}\naw{p^{(a)},q^{(b)}}}
 with $\rho_a\geq 0$, $\sigma_a\geq 0$, $\sum\limits_{a=1}^4\rho_a=\sum\limits_{a=1}^4\sigma_a=1$.
Now, in order to classify all Nash equilibria we have to find all sets $\naw{\sigma_a,p^{(a)}}$ and $\naw{\rho_b,q^{(b)}}$ such that $\$_A$ as a function of first strategy maximizes on $\naw{\sigma_a,p^{(a)}}$ while $\$_B$ as a function of second strategy maximizes on  $\naw{\rho_b,q^{(b)}}$. Let $\Lambda\subset\poisson{1,2,3,4}$ be the set of indices a such that $\sigma_a>0$. Then $\$_A\naw{\mu,\nu}$ is a convex combination of the quantities $\sum\limits_{b=1}^4\rho_b\$_A\naw{p^{(a)},q^{(b)}}$. Such a convex combination acquires a maximal value iff these quantities take the same (maximal) value for all $a\in\Lambda$. Consider now the payoff
\eq{\$_A\naw{p,\nu}\equiv\sum_{b=1}^4\rho_b\$_A\naw{p,q^{(b)}}.\label{c4}}
It is a quadratic form in $p$. Therefore, it takes the maximal values for $p$ belonging to the eigenspace of the relevant   quadratic matrix which corresponds to maximal eigenvalue. We conclude that the vectors $p^{(a)}$, $a\in\Lambda$, span this eigenspace; in other words, the highest eigenvalue has $\modu{\Lambda}$-fold degeneracy.

Similarly, let $\Sigma\subset\poisson{1,2,3,4}$ be the set of inidces $b$ such that $\rho_b>0$. Then the eigenspace of the matrix defined by the quadratic form in $q$
\eq{\$_B\naw{\mu,q}=\sum_{a=1}^4\sigma_a\$_B\naw{p^{(a)},q}\label{c5}}
which corresponds to the maximal eigenvalue is spanned by the vectors $q^{(b)}$, $b\in\Sigma$.

 Concluding, the search for Nash equilibria reduces to the problem of finding two sets $\mu=\poisson{\sigma_a,p^{(a)}}$ and $\nu=\poisson{\rho_b,q^{(b)}}$ such that $p^{(a)}$, $a\in\Lambda$ span the maximal eigenspace of the quadratic form $\$_A\naw{p,\nu}$, eq. (\ref{c4}), while $q^{(b)}$, $b\in\Sigma$ span the maximal eigenspace of quadratic form $\$_B\naw{\mu,q}$, eq. (\ref{c5}).
 
 The pair $\poisson{\poisson{\sigma_a,p^{(a)}},\poisson{\rho_b,q^{(b)}}}$ is called a Nash equilibrium of the $\naw{M,N}$ type  if $\modu{\Lambda}=M$, $\modu{\Sigma}=N$ (cf. \cite{Landsburg4}); obviously, $1\leq M,N\leq 4$ and, due to the symmetry of the game, we can assume $M\geq N$. 
 
 \section{Nash equilibria for the original ELW game}
 We will classify now the Nash equilibria for the original ELW game \cite{EisertWL} which corresponds to $X_0=3$, $X_1=5$, $X_2=0$, $X_3=1$.
 
 Let us start with $N=1$ case, i.e. the case Bob plays pure strategy. Due to the symmetry $p\rightarrow pr$, $q\rightarrow r^{-1}q$ one can assume Bob symmetry corresponds to $q=\mathbb{1}$. Then
 \eq{\$_A=\sum_{a=1}^4\rho_a\sum_{\alpha=0}^3X_\alpha\naw{p^{(a)}}_\alpha^2.}
 Note that $\rho_a\naw{p^{(a)}}_\alpha^2\geq 0$, $\sum\limits_{a=1}^4\sum\limits_{\alpha=0}^3\rho_a\naw{p^{(a)}}_\alpha^2=1$. Therefore, the maximum of $\$_A$ is achieved for $\rho_a=1$ for some $a$, say $a=1$, and $p^{(1)}=\pm e_1$. Then $\$_A=5$ and Alice plays pure strategy $p=e_1$. Consequently,
 \eq{\$_B\naw{\pm e_1,q}=\sum_{\alpha=0}^3\widetilde{X}_\alpha\naw{e_1q}_\alpha^2}
 and the maximum is achieved for $e_1q=\pm e_2$, i.e. $q=\pm e_3\neq e_0$; no Nash equilibrium exists.
 
 Assume now N=2. Due to our symmetry one can assume $q^{(1)}=\mathbb{1}=e_0$, $q^{(2)}=q$, $q=q_1e_1+q_2e_2+q_3e_3$, $q^2=-e_0$. Denote $\rho_1=\rho$, $\rho_2=1-\rho$, $0<\rho<1$. We find 
\eq{\$_A\naw{p,\nu}=\rho\sum_{\alpha=0}^3X_\alpha p_\alpha^2+\naw{1-\rho}\sum_{\alpha=0}^3\naw{pq}_\alpha^2.\label{c6}}
Let $X=\text{diag}\naw{X_0,X_1,X_2,X_3}$; moreover, we put (cf. eq. (\ref{c1}))
\eq{\naw{pq}_\alpha\equiv \widetilde{m}_{\alpha\beta}\naw{q}p_\beta.}
Then
\eq{\widetilde{m}\naw{q}=\left(\begin{array}{cccc}
q_0 & -q_1 & -q_2 & -q_3\\
q_1 & q_0 & q_3 & -q_2\\
q_2 & -q_3 & q_0 & q_1\\
q_3 & q_2 & -q_1 & q_0
\end{array}\right);}
$q^2=-e_0$, i.e. $q_0=0$, implies $\widetilde{m}^T\naw{q}=-\widetilde{m}\naw{q}$ and $\widetilde{m}^2\naw{q}=-I$. Eq. (\ref{c6}) can be rewritten as
\eq{\$_A\naw{p,\nu}=p^T\naw{\rho X+\naw{1-\rho}\widetilde{m}^T\naw{q}X\widetilde{m}\naw{q}}p.\label{d3}} 
According to the discussion of previous Section the matrix
\eq{Y\naw{q,\rho,X}\equiv\rho X+\naw{1-\rho}\widetilde{m}^T\naw{q}X\widetilde{m}\naw{q}\label{d4}} 
posseses $M$-fold ($M\geq 2$) degenerate highest eigenvalue.

$Y\naw{q,\rho,X}$ is a real symmetric matrix; explicitly
\eq{Y\naw{q,\rho,X}=\left(\begin{array}{cccc}
Y_0 & \naw{1-\rho}X_{32}q_2q_3 & \naw{1-\rho}X_{13}q_1q_3 & \naw{1-\rho}X_{21}q_1q_2\\
\naw{1-\rho}X_{32}q_2q_3 & Y_1 & \naw{1-\rho}X_{03}q_1q_2 & \naw{1-\rho} X_{02}q_1q_3\\
\naw{1-\rho}X_{13}q_1q_3 & \naw{1-\rho}X_{03}q_1q_2 & Y_2 & \naw{1-\rho}X_{01}q_2q_3\\
\naw{1-\rho}X_{21}q_1q_2 & \naw{1-\rho}X_{02}q_1q_3 & \naw{1-\rho}X_{01}q_2q_3 & Y_3
\end{array}\right)\label{d7}} 
where $X_{\alpha\beta}\equiv X_\alpha-X_\beta$ and
\eq{Y_0=\rho X_0+\naw{1-\rho}\naw{X_1q_1^2+X_2q_2^2+X_3q_3^2}}
\eq{Y_1=\rho X_1+\naw{1-\rho}\naw{X_0q_1^2+X_2q_3^2+X_3q_2^2}}
\eq{Y_2=\rho X_2+\naw{1-\rho}\naw{X_0q_2^2+X_1q_3^2+X_3q_1^2}}
\eq{Y_3=\rho X_3+\naw{1-\rho}\naw{X_0q_3^2+X_1q_2^2+X_2q_1^2}.}
Assume first that only one component of $q$ is nonzero. Then the matrix $Y\naw{q,\rho,X}$ is diagonal
\begin{footnotesize}
\eq{Y\naw{q,\rho,X}=\left(\begin{array}{cccc}
\rho X_0+\naw{1-\rho}X_3 & 0 & 0 & 0\\
0 & \rho X_1+\naw{1-\rho}X_2 & 0 & 0\\
0 & 0 & \rho X_2+\naw{1-\rho}X_1 & 0\\
0 & 0 & 0 & \rho X_3+\naw{1-\rho}X_0
\end{array}\right).}
\end{footnotesize}
Using the actual values of the payoffs $X_\alpha$ we find easily that the highest eigenvalue is degenerated only provided $\rho=\frac{1}{2}$. Then the eigenspace corresponding to the maximal eigenvalues is spanned by $e_1$ and $e_2$. Therefore, according to the general discussion presented above, Alice plays the strategies
\eq{\begin{split}
& p^{(1)}=e_1\cos\theta+e_2\sin\theta\\
& p^{(2)}=-e_1\sin\theta+e_2\cos\theta
\end{split}}
with the probabilities $\sigma$ and $\naw{1-\sigma}$, respectively. Then the quadratic matrix defining the Bob payoff reads
\begin{scriptsize}
\eq{Z\naw{p,\sigma,X}=\left(\begin{array}{cccc}
Z_0 & 0 & 0 & \naw{2\sigma-1}cs\naw{\widetilde{X}_1-\widetilde{X}_2}\\
0 & Z_1 & \naw{2\sigma-1}cs\naw{\widetilde{X}_0-\widetilde{X}_3} & 0\\
0 &  \naw{2\sigma-1}cs\naw{\widetilde{X}_0-\widetilde{X}_3} & Z_2 & 0\\
 \naw{2\sigma-1}cs\naw{\widetilde{X}_1-\widetilde{X}_2} & 0 & 0 & Z_3
 \end{array}\right)}
\end{scriptsize}
where $c\equiv\cos\theta$, $s\equiv\sin\theta$ and
\eq{Z_0=\widetilde{X}_1\naw{s^2+\sigma\naw{c^2-s^2}}+\widetilde{X}_2\naw{c^2-\sigma\naw{c^2-s^2}}\label{d1}}
\eq{Z_1=\widetilde{X}_0\naw{s^2+\sigma\naw{c^2-s^2}}+\widetilde{X}_3\naw{c^2-\sigma\naw{c^2-s^2}}}
\eq{Z_2=\widetilde{X}_0\naw{c^2-\sigma\naw{c^2-s^2}}+\widetilde{X}_3\naw{s^2+\sigma\naw{c^2-s^2}}}
 \eq{Z_3=\widetilde{X}_1\naw{c^2-\sigma\naw{c^2-s^2}}+\widetilde{X}_2\naw{s^2+\sigma\naw{c^2-s^2}}.\label{d2}}
 Now, $q=e_0$ should be an eigenvector corresponding to the highest eigenvalue. This implies
 \eq{\naw{2\sigma-1}\cos\theta\sin\theta=0.\label{d}}
If $\sigma=\frac{1}{2}$ we find double degeneracy of highest eigenvalue corresponding to $q^{(1)}=e_0$, $q^{(3)}=\pm e_3$. Therefore, we find the whole family of Nash equilibria parametrized by an angle $\theta$:
\begin{enumerate}
\item[] Alice:  
\begin{itemize}
\item[]The strategies
\eq{\begin{split}
& p^{(1)}=e_1\cos\theta+e_2\sin\theta\\
& p^{(2)}=-e_1\sin\theta+e_2\cos\theta
\end{split}\label{d5}}
played with the probabilities $\frac{1}{2}$;
\end{itemize}
\item[]Bob:
\begin{itemize}
\item[] The strategies
\eq{\begin{split}
& q^{(1)}=e_0\\
& q^{(2)}=\pm e_3
\end{split}\label{d6}}
played with the probabilities $\frac{1}{2}$.
\end{itemize}
\end{enumerate}
Let us note that both the sign arbitrariness and the arbitrary value of the $\theta$ angle are the consequences of the $O(d)$ ($d=2$) symmetry present in the case of degeneracy. 
The second solution to the equation (\ref{d}) reads
\eq{\cos\theta\sin\theta=0.}
Using actual values of $\widetilde{X}_\alpha'$we find easily from eqs. (\ref{d1})$\div$(\ref{d2}) that $Z$ has degenerate highest eigenvalue only provided $\sigma=\frac{1}{2}$ and the second eigenvector corresponding to this eigenvalue is $q=\pm e_3$.

Summarizing, we have found that, if $N=2$ and we assume that the vector $q$ entering eqs. (\ref{d3}) and (\ref{d4}) has only one nonvanishing component the only set of Nash equilibria is given by eqs. (\ref{d5}) and (\ref{d6}).

The case $N=2$ with at least two nonvanishing components of $q$ is more involved. let $T^{(\alpha)}\subset \mathbb{R}^4$ be the subspace defined by the relations
\eq{v\in T^{(\alpha)}\quad \text{iff} \quad v_\alpha=0,\quad \naw{Y\naw{q,\rho,X}v}_\alpha=0}  
where the matrix $Y\naw{q,\rho,X}$ is given by eq. (\ref{d7}). Under our assumption concerning $q$ the spaces $T^{(\alpha)}$ are twodimensional for $\alpha=0,1,2,3$.\\
Defining
\begin{equation}
u^{(0)}=\left(\begin{array}{c}
0\\
q_1\\
q_2\\
q_3
\end{array}\right),\qquad
v^{(0)}=\left(\begin{array}{c}
0\\
X_1q_1\\
X_2q_2\\
X_3q_3
\end{array}\right)\label{abc}
\end{equation}
 \begin{equation}
u^{(1)}=\left(\begin{array}{c}
q_1\\
0\\
q_3\\
-q_2
\end{array}\right),\qquad
v^{(1)}=\left(\begin{array}{c}
X_0q_1\\
0\\
X_2q_3\\
-X_3q_2
\end{array}\right)
\end{equation}
 \begin{equation}
u^{(2)}=\left(\begin{array}{c}
q_2\\
-q_3\\
0\\
q_1
\end{array}\right),\qquad
v^{(2)}=\left(\begin{array}{c}
X_0q_2\\
-X_1q_3\\
0\\
X_3q_1
\end{array}\right)
\end{equation} 
\begin{equation}
u^{(3)}=\left(\begin{array}{c}
q_3\\
q_2\\
-q_1\\
0
\end{array}\right),\qquad
v^{(3)}=\left(\begin{array}{c}
X_0q_3\\
X_1q_2\\
-X_2q_1\\
0
\end{array}\right)
\end{equation}
we easily find that $u^{(\alpha)},\, v^{(\alpha)}\in T^{(\alpha)}$; moreover, $u^{(\alpha)}$ and $ v^{(\alpha)}$ span $T^{(\alpha)}$. Assume that the highest eigenvalue of $Y\naw{q,\rho,X}$ is at least doubly degenerated. Then there exists an eigenvector $w$ of $Y\naw{q,\rho,X}$ which obeys (say) $w_0=0$,
\begin{equation}
Yw=\lambda w\label{d9}
\end{equation} 
Eq. (\ref{d9}) implies that $w\in T^{(0)}$. By virtue of this one can write 
\begin{equation}
w=au^{(0)}+bv^{(0)},\quad a,b\in\mathbb{R}\label{d10}
\end{equation}
Inserting the general form of $w$, as given by eq. (\ref{d10}) into eq. (\ref{d9}) we find, using 	MATHEMATICA, that eq. (\ref{d9}) is fulfiled if either (a) $q_1=0$ or $q_2=0$ or (b) $\rho=\frac{1}{2}$.\\
Consider the case (a). Let us first assume that $q_1=0$. The matrix $Y\naw{q,\rho,X}$ simplifies considerably; only two nondiagonal elements are nonvanishing. $Y\naw{q,\rho,X}$ can be diagonalized explicitly (the best way is to use MATHEMATICA). Imposing the condition that the largest eigenvalue of $Y\naw{q,\rho,X}$ is doubly degenerate we find $q_2$ nad $q_3$ as the functions of $\rho$. It appears that for $\rho=\frac{1}{2}$ we recover our old solution (\ref{d5}), (\ref{d6}). For $\rho\neq\frac{1}{2}$ there is a one parameter family of solutions $q_2=q_2(\rho)$, $q_3=q_3(\rho)$; $0\leq q_2^2\leq 1$ and $0\leq q_3^2\leq 1$ imply $\frac{1}{3}\leq\rho\leq\frac{2}{3}$. Denoting again by $\tilde{p}^{(1)}$ and $\tilde{p}^{(2)}$ the eigenvectors of $Y\naw{q,\rho,X}$ corresponding to the largest eigenvalue we find that the Alice strategy is
\begin{equation}
\begin{split}
& p^{(1)}=\tilde{p}^{(1)}\cos\theta+\tilde{p}^{(2)}\sin\theta\\
& p^{(2)}=-\tilde{p}^{(1)}\sin\theta+\tilde{p}^{(2)}\cos\theta\\
\end{split}
\end{equation} 
played with the probabilities $\sigma$ and $\naw{1-\sigma}$, respectively. Now, we demand that the Bob payoff maximizes on $e_0$ and $q$. There are three free parameters to be determined: $\theta$, $\sigma\in\naw{0,1}$ and $\rho\in\av{\frac{1}{3},\frac{2}{3}}$ ($\rho\neq\frac{1}{2}$). Straightforward computations with the help of 		MATHEMATICA	 shows that no solutions exists. Similar reasoning shows that there is no solution also if $q_2=0$.

Let us now consider the case (b), $\rho=\frac{1}{2}$. The matrix $Y$ reads then (cf. eq. (\ref{d4}))
\begin{equation}
Y\naw{q,\frac{1}{2},X}=\frac{1}{2}\naw{X-\widetilde{m}(q)X\widetilde{m}(q).}
\end{equation}
Note that then
\begin{equation}
Y\naw{q,\frac{1}{2},X}\widetilde{m}\naw{q}=\widetilde{m}\naw{q}Y\naw{q,\frac{1}{2},X}.\label{cd}
\end{equation}
Therefore, if $v$ is an eigenvector of $Y$ then $\widetilde{m}(q)v$ is also an eigenvector corresponding to the same eigenvalue. Moreover, due to the antisymmetry of $\widetilde{m}(q)$, $\widetilde{m}(q)v$ is orthogonal to $v$ (and nonvanishing as $\widetilde{m}^2(q)=-I$). So the eigenvalues of $Y$ are doubly degenerated. There are two possibilities: either all four eigenvalues are equal or there are two eigenvalues, $\lambda_1>\lambda_2$, each doubly degenerate. In the first case $Y=\lambda I$ and at least two component of $q$ must vanish contrary to our assumption. We conclude that $Y\naw{q,\frac{1}{2},X}$ has two doubly degenerated eigenvalues for any $q$ under consideration. Therefore, $M=2$; exchanging the players and making use of the symmetry of the game we conclude the previous arguments, applied now to Alice strategy, imply $\sigma=\frac{1}{2}$. Alice plays using, with the same probability, two orthogonal strategies $p^{(1)}$ and $p^{(2)}$ which are the eigenvectors of $Y\naw{q,\frac{1}{2},X}$ corresponding to the highest eigenvalue. Again, by virtue of eq. (\ref{cd}) $\widetilde{m}(q)p^{(1)}$ is a unit eigenvector of $Y\naw{q,\frac{1}{2},X}$ orthogonal to $p^{(1)}$ so we have
\begin{equation}
p^{(2)}=\pm\widetilde{m}\naw{q}p^{(1)}.\label{cd1}
\end{equation}
The quadratic matrix corresponding to the Bob payoff can be written in the form
\begin{equation}
Z\naw{\underline{p},\frac{1}{2},\widetilde{X}}=m^T\naw{p^{(1)}}\widetilde{X}m\naw{p^{(1)}}+m^T\naw{p^{(2)}}\widetilde{X}m\naw{p^{(2)}}
\end{equation}
or, using eq. (\ref{cd1})
\begin{equation}
Z\naw{\underline{p},\frac{1}{2},\widetilde{X}}=m^T\naw{p^{(1)}}\widetilde{X}m\naw{p^{(1)}}+m^T(q)m^T\naw{p^{(1)}}\widetilde{X}m\naw{p^{(1)}}m(q).\label{cd5}
\end{equation}
We see that, in analogy to eq. (\ref{cd})
\begin{equation}
Z\naw{\underline{p},\frac{1}{2},\widetilde{X}}m(q)=m(q)Z\naw{\underline{p},\frac{1}{2},\widetilde{X}}.\label{cd3}
\end{equation}
Now, $e_0$ must be the eigenvector of $Z$ corresponding to the highest eigenvalue. By virtue of eq. $q=m(q)e_0$  (\ref{cd3}) $q$ is automatically also such an eigenvector. We conclude that in order to find the Nash equilibrium for general $q$ one has to select some eigenvector $p^{(1)}$, corresponding to highest eigenvalue of $Y\naw{q,\frac{1}{2},X}$ and to define $p^{(2)}$ by eq. (\ref{cd1}). Then one should check if $e_0$ is an eigenvector of $Z\naw{p,\frac{1}{2},\widetilde{X}}$ given by eq. (\ref{cd5}), again corresponding to its highest eigenvalue. Let us note that the $O(2)$ symmetry due to $\rho=\frac{1}{2}$ allow us to select $p^{(1)}$ in such a way that one of its component (for example, $p_0^{(1)}$) vanishes. It is not difficult to check then using MATHEMATICA that no solution exists unless two components of $q$ vanish. This completes the case $M=N=2$.

Let us note that the cases $N=2$, $M=3,4$ can solved quite easily. If $M=3$ the maximal eigenvalue of $Y\naw{q,\rho,X}$ has a triple degeneracy. Then one can choose at least two linearly independent eigenvectors $w^{(1)}$, $w^{(2)}$ belonging to (say) $T^{(0)}$. So they span $T^{(0)}$ and, consequently, $u^{(0)}$ and $v^{(0)}$ (cf. eq. (\ref{abc})) are eigenvectors  of $y\naw{q,\rho,X}$. It is easy to check that it is impossible for generic $q$. The case $M=4$ yields $Y\naw{q,\rho,X}=\lambda I$ which implies that only one component of $q$ is nonvanishing.

Finally, consider the case $N\geq 3$. One can again write out the relevant matrices $Y$ and $Z$ and derive the conditions to be fulfield in order they posses at least threefold degenerate maximal highest eigenvalues. However, it is more convenient to use directly the results derived by Landsburg \cite{Landsburg1} which take particularly simple form when $M\geq N\geq 3$. Namely, there are two possibilities: either (i) each players strategy is supported on three of the four strategies $\pm e_0,\, \pm e_1,\, \pm e_2,\, \pm e_3$ or (ii) each player plays each of four orthogonal quaternions with probability $\frac{1}{4}$. It is straightforward to check that for the actual values of payoffs $X_\alpha$ the case (i) leads to no Nash equilibria. As far as (ii) is concerned, the $O(4)$ degeneracy allows us to assume that both player play strategies supported on $e_0,\, e_1,\, e_2$ and $e_3$. Again no Nash equilibrium is attained.

\section{Explicit form of Nash equilibria}
The following conclusion follows from the results derived in previous sections. All Nash equilibria for the original maximally entangled ELW game are obtained from the following strategies of Alice and Bob
\begin{enumerate}
\item[] Alice:
\begin{itemize}
\item[] $e_1$ and $e_2$ played with equal probabilities
\end{itemize}
\item[] Bob:
\begin{itemize}
\item[] $e_0$ and $e_3$ played with equal probabilities,
\end{itemize}
\end{enumerate}
by applying the following operations: 
\begin{itemize}
\item[(i)] $O(2)$ transformations on $\naw{e_1,e_2}$ and $\naw{e_0,e_3}$
\item[(ii)] symmetry transformations $p\rightarrow pr$, $q\rightarrow r^{-1}q$ with $r$ being an arbitrary unit quaternion.
\end{itemize}
As a result we obtain the following family of strategies:
\begin{enumerate}
\item[] Alice:
\begin{itemize}
\item[] $\naw{e_1\cos\theta_A+e_2\sin\theta_A}r$, $\pm\naw{-e_1\sin\theta_A+e_2\cos\theta_A}r$\\ played with equal probabilities 
\end{itemize}
\item[] Bob:
\begin{itemize}
\item[] $r^{-1}\naw{e_0\cos\theta_B+e_3\sin\theta_B}$, $\pm r^{-1}\naw{-e_0\sin\theta_B+e_3\cos\theta_B}$\\  played with equal probabilities.
\end{itemize}
\end{enumerate}
One can get rid of $\theta_B$ dependence by factorizing an arbitrary $r$ as $r(\theta_B)\cdot r$
with
\eq{r(\theta_B)=e_0\cos\theta_B+e_3\sin\theta_B.}
Finally, we arrive at the following form of strategies:
\begin{enumerate}
\item[] Alice:
\begin{itemize}
\item[] $\naw{e_1\cos\theta+e_2\sin\theta}r$, $\pm\naw{-e_1\sin\theta+e_2\cos\theta}r$\\ 
\end{itemize}
\item[] Bob:
\begin{itemize}
\item[] $r^{-1}e_0$, $\pm r^{-1}e_3$
\end{itemize}
\end{enumerate}  
played with equal probabilities.

Let us translate our result back to the language of $SU(2)$ matrices. Taking into account eqs. (\ref{z}) and (\ref{z1}) and keeping in mind that we have made the replacement $Q\rightarrow q^{-1}$ we find the following form of Nash strategies:
\begin{enumerate}
\item[] Alice:
\begin{equation}
U_A^{(1)}=\left(\begin{array}{cc}
-\beta e^{-i\theta} &  -i\alpha e^{-i\theta}\\
-i\overline{\alpha}e^{i\theta} & -\overline{\beta}e^{i\theta}\end{array}\right),\qquad U_A^{(2)}=\left(\begin{array}{cc}
i\beta e^{-i\theta} &  -\alpha e^{-i\theta}\\
\overline{\alpha}e^{i\theta} & -i\overline{\beta}e^{i\theta}\end{array}\right)
\end{equation}
\item[] Bob:
\begin{equation}
U_B^{(1)}=\left(\begin{array}{cc}
\alpha &  \beta\\
-\overline{\beta} & \overline{\alpha}\end{array}\right),\qquad U_B^{(2)}=\left(\begin{array}{cc}
-i\alpha &  -i\beta\\
-i\overline{\beta} & i\overline{\alpha}\end{array}\right)
\end{equation}
\end{enumerate} 
all strategies being played with probability $\frac{1}{2}$.\\
Let us note that the substitution 
\begin{equation}
\begin{split}
& \alpha'=-\beta e^{-i\theta}\\
& \beta'=-i\alpha e^{-i\theta}\\
& \theta'=\pi-\theta
\end{split}
\end{equation}
exchanges the roles of Alice and Bob. Puting $\alpha=0$, $\beta=1$ and $\theta=\pi$ yields the Nash equilibrium considered in Ref. \cite{EisertW}.

\subsection*{Acknowledgment}
Research of Katarzyna Bolonek-Laso\'n was supported by NCN Grant\\ no. DEC-2012/05/D/ST2/00754.

\end{document}